\documentclass[twocolumn,showpacs,prl]{revtex4}
\usepackage{graphicx}
\usepackage{amsmath}

\newcommand{\bi}{\begin{itemize}}
\newcommand{\ei}{\end{itemize}}
\newcommand{\bR}{{\bf R}}

\newcommand{\ie}{{\it i.e., }}

\begin{document}

\title{
{\it Published in Advances in Quantum Monte Carlo, Editor(s): Shigenori Tanaka, Stuart M. Rothstein, William A. Lester , Jr., ACS Symposium Series, Volume 1094, American Chemical Society, (peer rev.),  pp 77-87 (2012)}
 \break
\qquad \break
\qquad
Many-body nodal hypersurface and domain averages for correlated 
wave functions} 

\author{Shuming Hu, Kevin Rasch, Lubos Mitas}
\affiliation{
Department of Physics,
North Carolina State University, Raleigh, NC 27695.}

\date{\today}

\begin{abstract}
We outline the basic notions of nodal hypersurface and domain averages for
antisymmetric wave functions. We illustrate their properties and analyze the 
results for a few 
electron explicitly solvable cases and discuss possible further developments. 
\end{abstract}

\maketitle

Quantum Monte Carlo is  one of the most effective many-body methodologies
for the study of quantum
systems. It is based on
a combination of analytical insights, robustness of stochastic approaches,
and performance of parallel architectures [1-10]. The approach has been
applied to a variety of challenging problems in electronic structure
of atoms, small molecules, clusters, solids, ultracold condensates,
and beyond [1-15]. 
The two most commonly used QMC methods are variational Monte Carlo and
diffusion Monte Carlo (DMC). Let us briefly recapitulate the basics of the DMC method.

It is straightforward to show that for $\tau \to \infty$,
 the operator $\exp(-\tau H)$
  projects out the ground state of a given symmetry
 from any trial function with nonzero overlap.  We assume that the  
Hamiltonian $H$ is time-reversal symmetric 
so that the eigenstates can be chosen to be real.  This projection  is most
 conveniently carried out by solving the Schr\"odinger equation in
 an imaginary time integral form so that the product $f(\bR,t)=$ $\Phi(\bR,t)\Psi_T(\bR)$
 obeys
\begin{equation*}
f(\bR,t+\tau)=\int G(\bR,\bR',\tau)f(\bR',t)\text{d}\bR',
\end{equation*}
where 
\begin{equation*}
f(\bR,t)= \Phi(\bR,t)\Psi_T(\bR).
\end{equation*}
The Green's function is given by
\begin{equation*}
G(\bR,\bR',\tau)=\frac{\Psi_T(\bR)}{\Psi_T(\bR')}
\langle \bR|\exp[-\tau(H-E_T)]|\bR' \rangle,
\end{equation*}
where 
${\bf R}=({\bf r}_1, ..., {\bf r}_N)$
denotes positions of $N$ particles
and $E_T$ is an energy offset.
 In the DMC method,
 the function $f(\bR,t)$ is represented by a set of 10$^2$-10$^4$
 random walkers (sampling points) in the $3N$-dimensional space of
 electron configurations. The walkers are propagated for
 a time slice $\tau$ by interpreting the Green's function
 as a transition probability from $\bR'$ $\to$
 $\bR$. The kernel is known
 for small $\tau$, and the large time $t$
 limit is obtained
 by iterating the propagation.  The method is formally exact provided
 that the boundary conditions, \ie   the {\em fermion nodes}  of
 the antisymmetric solution defined as $\Phi(\bR,\infty )=0$,
 are known [1,5,14].
 Unfortunately the antisymmetry does not specify the nodes completely,
 and currently we have to use approximations.  The
 commonly used fixed-node approximation [14]
 enforces the nodes of $f(\bR,t)$ to
 be identical to the nodes of $\Psi_T(\bR)$ which then 
implies that
\begin{equation*}
f(\bR,t)\ge 0
\end{equation*}
everywhere.
It is therefore clear that the accuracy of the fixed-node DMC is determined by the quality of 
the trial wave function nodes.
The commonly used form for $\Psi_T$ is the Slater-Jastrow wave function given as
\begin{equation*}
\Psi_T(\bR) =
\sum_n d_n 
\text{det}^{\uparrow}_n\left[\{\varphi_{\alpha}\}\right]
\text{det}^{\downarrow}_n\left[\{\varphi_{\beta}\}\right]
\exp\left[U_{corr}\right]
\end{equation*}
 where $U_{corr}$ is the correlation factor explicitly depending 
on interparticle distances thus describing pair or higher order
correlations explicitly.
 The typical number of Slater
 determinants is between 1 and 10$^3$,
 and the corresponding weights $\{d_n\}$ are usually
 estimated
 in multi-reference Hartree-Fock (HF) or Configuration Interaction
 (CI)
 calculations and then re-optimized in the variational framework.

 It is quite remarkable that the nodes of such Slater-Jastrow wave functions  
(often with a single-determinant
product only) lead to unexpectedly small errors 
 and that the typical amount of obtained correlation
 energy in fixed-node DMC is $\approx$ 95\%.
 This is true for essentially all systems
 we have studied: atoms, molecules, clusters and solids [1-15].

The fixed-node approximation 
 is perhaps the single most important 
unsolved problem which hampers the progress in further improvement of
 accuracy and efficiency of the QMC calculations. 
One of the key difficulties is that the fixed-node bias is actually very small
on the scale of the total energy.  A comparison of the total energy components
for a typical electronic structure problem is given in Table I.

\begin{table}[h]
\caption{Energy components as percentages of the total energy
in Coulombic systems.
$\Delta_{FN}=E_{exact}-E_{FNDMC/HF}$ is the  fixed-node (FN)  bias 
corresponding to the Hartree-Fock nodes.}
\label{tab:comparison3}
\begin{tabular}{ll}
\hline
Energy component\quad \quad & \% of E$_{tot}$  \\
\hline
Kinetic & 100  \\
Exchange &   $\approx$ 8 \\ 
Correlation &   $\approx$ 3 \\ 
$\Delta_{FN}=E_{exact}-E_{FNDMC/HF}$ \qquad &  $\approx$ 0.2  \\
\hline
\end{tabular}
\end{table}
 
Considering the typical fluctuations of the DMC energy per stochastic 
sample (which is of the order of a few percent of the total energy), the node-related
``signal"  is very weak. 
Unfortunately a few percent of the correlation energy can influence the energy 
 differences we are interested in.

The optimization methods (despite
a number of recent developments) have difficulties picking up the nodal bias 
signal as it appears buried in the noise which is inherent to the QMC 
methodology. Recent developments in the nodal optimization
using the self-healing method [16] enables to filter out some of the 
noise quite effectively, however, the performance of the method has to be
tested on more systems.  
  However, this is not the only problem. Another key issue is that
our knowledge of the nodal properties is very limited.  At present we simply have 
no clear idea how to improve the nodal hypersurfaces for general cases
in an efficient manner (for example, in systems which might require
an exponential number of Slater determinants just to 
describe the correct spin and spatial symmetries).

We simply have to develop other measures  which can provide more targeted information
about the nodal shapes. It is straightforward to show that 
the total energy or its components are not  selective enough in this respect. Let us consider a few 
simple illustrations.
For example, the two non-interacting two-electron atomic states
$^3S(1s2s)$ and $^3P(1s2p)$ have the same total, kinetic, and potential energies,
but different nodal shapes.
Since the symmetries in this case are different, one might argue that the symmetry 
should be used to distinguish and, possibly,
classify the nodal shapes in this case. Consider another case:  
 non-interacting 
four-electron atomic states $^1S(1s^22s^2)$ and $^1S(1s^22p^2)$.
These degenerate states have the same
symmetry, however, the nodes are different, both in topology and in the shape. 
Clearly, we would like to 
measure and distinguish the nodes in such cases. 

Characterization of the nodes can be of significant interest in another context.
Very recently, very interesting scenario was suggested for systems in  
quantum critical point, namely, that 
the nodes in such state might exhibit fractal (scale-invariant)
 character [16]. For this purpose it would be very useful to measure 
the smoothness of the nodal surfaces.

Let us now consider the stationary Schr\"odinger equation
\begin{equation*}
-(1/2)\nabla_{\bf R}^2\Psi({\bf R}) +V({\bf R})\Psi({\bf R})=E\Psi({\bf R}),
\end{equation*}
where $({\bf R})=({\bf r}_1,{\bf r}_2, ..., {\bf r}_N)$. The exact fermionic 
eigenstate $\Psi$
determines the nodal domains
\begin{equation*}
\Omega^+=\{{\bf R}; \Psi({\bf R})>0\}, \; \Omega^-=\{{\bf R}; \Psi({\bf R})<0\}
\end{equation*}
and the corresponding node $\partial\Omega$. We integrate the 
equation over the $\Omega^+$ domain {\em } only, and using the
 Gauss-Stokes-Green theorem we get
\begin{equation*}
\int_{\Omega^+} (V({\bf R})-E)\Psi({\bf R}) \, \text{d}{\bf R}
-\frac{1}{2}\int_{\partial\Omega}\nabla_{\bf R}\Psi({\bf R})\cdot \text{d}{\bf S}_n 
 =0.
\end{equation*}
Similarly, we can integrate over the  $\Omega^-$ domain and if we put it together
(assuming either free or periodic boundary conditions) then we get
\begin{equation*}
\int_{\partial\Omega}|\nabla_{\bf R}\Psi({\bf R})| \text{d}{\bf S}
+\int V({\bf R})|\Psi({\bf R})| \text{d}{\bf R} 
=E\int |\Psi({\bf R})| \text{d}{\bf R}.
\end{equation*}
The obtained equation shows that the total energy
is given as a sum of kinetic and potential components, which we call nodal (hypersurface) and domain 
averages (nda, in short). They are defined as follows
$$
E_{kin}^{nda}=\int_{\partial\Omega}|\nabla_{\bf R}\Psi({\bf R})|\text{d}{\bf S}/
\int |\Psi({\bf R})| \text{d}{\bf R},
$$
and
$$
E_{pot}^{nda}=\int V({\bf R})|\Psi({\bf R})|  \text{d}{\bf R}/
\int |\Psi({\bf R})| \text{d}{\bf R},
$$
 so that 
$$E=E_{kin}^{nda}+E_{pot}^{nda}.$$
This derivation and definitions 
 deserve some comments. First, we tacitly assumed that there is only one
positive and one negative nodal domain, however, this generically applies 
only to  
fermionic ground states.
Generalization to more domains is straightforward: one integrates domain
by domain and sums the results. It therefore applies to {\em any eigenstate} 
including excitations, both fermionic and bosonic (the bosonic ground state
is an exception since it is nodeless). 
It is important that $E_{kin}^{nda}$ depends solely on the gradient of the wave function
on the node (domain boundary) and not on the wave function values  inside the domain.
The key idea is that these expressions measure properties of the quantum
amplitudes more directly than the expectation values. In fact,
the expectations {\em supress} the nodal signal since both 
the square of the exact eigenstate and also its Laplacian vanish at the node. 
Note that although the sum of 
kinetic and potential nda components produces the total energy, the 
expression has no variational property, \ie it is not quadratic in the 
wave functions as is the usual expectation value. It is rather a ``one-sided
expectation'' which enables one to probe the nodal structure as we will show in what follows.

The nda values are not trivial to calculate, and for illustration we 
will present 
just a few simple
cases. Let us first consider a toy model,
 an electron in $2p$ orbital so that the state is $^2P(2p)$.
For the Coulomb potential $V(r)=-Z/r$ we have 
$$\Psi=z\rho_{2p}(r)=z\exp(-Zr/2)$$ 
and we can write
$$
E_{pot}^{nda}(2p) = \frac{
\int V(r)\rho_{2p}(r)r|\cos(\vartheta)|r^2\sin(\vartheta)\text{d}\vartheta \text{d}r}
{\int \rho_{2p}(r)r|\cos(\vartheta)|r^2\sin(\vartheta)\text{d}\vartheta \text{d}r} = 
\frac{-Z^2}{6}.
$$
The node is the plane given by $z=0$, and we can easily evaluate 
also the kinetic energy part since 
$$
|\nabla \Psi|_{z=0}=\rho_{2p}(r)
$$
so that we can write
$$
E_{kin}^{nda}(2p) = \frac{
\int \rho_{2p}(r)\text{d}x\text{d}y}{2\pi 
\int \rho_{2p}(r)r|\cos(\vartheta)|r^2\sin(\vartheta)\text{d}\vartheta \text{d}r}
=\frac{Z^2}{24}.
$$
Note that the integral in the numerator is over the plane while  
the integration domain of the denominator is the full 3D volume.  One can also
verify that the sum of the two components gives $E=-Z^2/8$ as expected. 

Much more interesting are cases with more than one particle.
We mentioned  the two excitations of the He atom, namely 
 $^3S(1s2s)$ and  $^3P(1s2p)$,
and also the corresponding
four-particle singlets  $^1S(1s^22s^2)$ and
 $^1S(1s^22p^2)$. Actually, these are quite 
nontrivial to calculate even in noninteracting cases.
The state $^3S(1s2s)$ is straightforward but rather involved,
and one ends up with numerous
integrals. The states with $2p$ orbitals are even more complicated
since the node is given by a combination of exponentials and linear 
functions so that the integration domains become complicated. Therefore
for this case we have used Monte Carlo integration. 
The resulting values are listed in Table II.

%

\begin{table}[h]\renewcommand{\arraystretch}{2}\addtolength{\tabcolsep}{-1pt}
\caption{Energy components for two- and four-electron atoms: standard 
expectations and nda values. The energies in a.u. are proportional to $Z^2$. The results are exact except for the values with
the error bars $\varepsilon\approx 1.10^{-5}$ in the brackets. The dot means that the value 
is the same as in the row above. ({\it NB: After this paper was published
we were able to get the analytical result for the $^3P(1s2p)$ state 
and we found that the given values are exact, 
ie, $\varepsilon$ below is zero.})}
\label{tab:comparison3}
\begin{tabular}{lccccc}
\hline
\hline
State  & \quad E$_{tot}$ \quad &\quad E$_{kin}$\quad & \quad E$_{pot}$ \quad & \quad E$_{kin}^{nda}$ & E$_{pot}^{nda}$ \\ [.1cm]
\hline
$^3S(1s2s)$ & -$\frac{5}{8}$  & $\frac{5}{8}$ & $-\frac{5}{4}$ 
& $\frac{10}{221}$ & $- \frac{1185}{1768}$ \\
$^3P(1s2p)$ & .  & . & . &\quad $\frac{1}{20} (\varepsilon) $ & \quad $-\frac{27}{40} (\varepsilon)$  \\
\hline
 $^1S(1s^22s^2)$ \quad & -$\frac{5}{4}$  & $\frac{5}{4}$ & $-\frac{5}{2}$ 
& $\frac{20}{221}$ & $- \frac{1185}{884}$ \\
 $^1S(1s^22p^2)$ \quad & .  & . & . & \quad $\frac{1}{10} (\varepsilon) $  &\quad $-\frac{27}{20} (\varepsilon)$ \\
\hline
\hline
\end{tabular}
\end{table}

The values show clearly that one can distinguish the states 
and the nodes by the nodal and domain averages. For example,
the $E_{kin}^{nda}$ differ by more than 10 (0.002)\%  between the corresponding degenerate rows. 
Note that if one would consider the interaction, then the two four-electron states would mix.  
Clearly, the nda components will depend on the mixing and thus reflect
the node change under interactions. If fact, there is an optimal mixing 
which provides the best node within the functional form as shown previously
in calculations of the Be atom [17,18].

It is interesting to analyze another case: two noninteracting electrons in $2p^2$ configuration
which can couple into the three states 
$^3\negthinspace P, ^1\negthinspace S, ^1\negthinspace D$.
 For example,
for the state $^3\negthinspace P(2p^2)$ the wave functions is given by
$$
\Psi(1,2) = \rho_{2p}(r_1)\rho_{2p}(r_2)(x_1y_2-x_2y_1),
$$
and the nda potential energy can be written as
$$
E_{pot}^{nda}(2p^2) = 2
\frac{\int V(r_1)\rho_{2p}(r_1)\rho_{2p}(r_2)|x_1y_2-x_2y_1|\text{d}{\bf r}_1 \text{d}{\bf r}_2}
{\int \rho_{2p}(r_1)\rho_{2p}(r_2)|x_1y_2-x_2y_1|\text{d}{\bf r}_1 \text{d}{\bf r}_2}.
$$
The integrals can be factorized into
radial and angular components.  Since the angular parts
cancel out, we get
$$
E_{pot}^{nda}(2p^2) = 2E_{pot}^{nda}(2p).
$$
It is perhaps somewhat unexpected that we get the same result also for
the other two states $^1\negthinspace S(2p^2)$
and
$ ^1\negthinspace D(2p^2)$.

\begin{table}[h]\renewcommand{\arraystretch}{2}\addtolength{\tabcolsep}{-1pt}
\caption{Energy components for $2p^2$ states for Coulomb potential: standard expectations 
and nda values}
\label{tab:comparison3}
\begin{tabular}{lccccc}
\hline
State  & E$_{tot}$ \qquad & E$_{kin}$ \qquad & E$_{pot}$\qquad & 
E$_{kin}^{nda}$ \qquad & E$_{pot}^{nda}$ \qquad \\
\hline
$^3\negthinspace P, ^1\negthinspace S, ^1\negthinspace D$ \quad \quad 
 & $\quad-\frac{1}{4}\quad$  & $\quad\frac{1}{4}\quad$ 
 & $\quad -\frac{1}{2}\quad $ & $\quad\frac{1}{12}\quad$ & $\quad-\frac{1}{3}\quad $ \\
\hline
\end{tabular}
\end{table}
We therefore conclude that all the components are the same for all three states,
although two of them are singlets and one is triplet and also have different
spatial symmetries. Note that this is strictly true only 
for the {\it noninteracting } system.
This implies that the states might have equivalent
nodes, and a little bit of analysis actually shows that. 
One can find that the node for the $^3P$ state can be described from the 
perspective of one of the two electrons as the plane defined by the angular momentum 
axis and the second electron. Similarly, the node of the $^1D$ state
looks to one of the electrons as the plane which contains the angular 
momentum axis and is orthogonal to a plane defined by the second electron
and the angular momentum axis. Finally, for $^1S$ states one of the electrons 
sees a plane which is orthogonal to the position vector of the second 
electron. In all three cases the node subset is therefore a plane 
which passes through the origin. Although these are only subsets of the complete
5D node, which is a hyperbolic hypersurface in 6D,  the construction
 enables us to get
an insight into their properties. In fact, this shows that
there are only two nodal domains in all three cases: the scanning electron is either
on one or the other side of the considered plane. Let us further 
 define the equivalency for a set of
 nodes. By the equivalency we mean that the nodes in the given set can be transformed
to each other by coordinate transforms 
 which are unitary (the determinant of the  transformation matrix is equal to
+1 or -1). This includes not only rotations but also reflections around the 
origin since otherwise the triplet nodes cannot be transformed to the singlet nodes. 
This can be inspected, for example, by transforming the node of one
of the $^1D$ states 
$$
\Psi(1,2) = \rho_{2p}(r_1)\rho_{2p}(r_2)(x_1y_2+x_2y_1),
$$
to the node of $^3P$ state using the reflection of one 
coordinate component, say, $x_2 \to -x_2$ (see the wave function above).
With some effort one could find that the nodes of the singlets are also 
equivalent. For this the reader might find
useful to consult our previous papers on related topics of nodal structure
and analysis [9].

Note that for {\em interacting}
two electrons in these states 
the nda components will {\it not} be identical since the e-e Coulomb repulsion will
distort the wave function gradients
in different ways for different states, and energetically,
it will favor the triplet over the singlets.

The previous case of two non-interacting electrons can be further generalized 
to a given subshell $l=n-1$ for any $n$ and for any
possible spin symmetry and occupations up to
the maximum $2(2l+1)$. The energies can be evaluated 
the same way as above, and it is revealing
 to explore 
 the quasiclassical limit of the nda estimates. 
Consider the class of atomic (excited) states such that $k$ electrons occupy
subshell $l=n-1$ with any allowed spatial and spin symmetry.
The state is $^{2S+1}L[\varphi_l^k]$ where $k$ is the occupation.
One can find:
$$
E_{kin}^{nda}(k,l)=kZ^2 \frac{l}{2(l+1)^2(l+2)}
$$
and
$$
E_{pot}^{nda}(k,l)=-kZ^2 \frac{1}{(l+1)(l+2)}
$$
so that all the non-interacting nodes for various symmetries are equivalent.
By checking out the quasiclassical limit $l\to \infty$,
we find
$$
\lim_{l\to\infty} E_{kin}^{nda}(k,l)=\lim_{l\to\infty} E_{kin}(k,l)
$$
and
$$
\lim_{l\to\infty} E_{pot}^{nda}(k,l)=\lim_{l\to\infty} E_{pot}(k,l).
$$
Clearly, averages over $\Psi^2$ and $|\Psi|$ become identical since the quantum
effects become irrelevant for  $\lim_{l\to\infty}$.

Let us  now turn to the case of a system with interactions. Consider the 
two-particle 3D harmonic problem with the Coulomb  interaction. The Hamiltonian
is given by 
$$
H=(P_1^2 +P_2^2)/2 + \omega^2(r_1^2+r_2^2)/2 +g_0/r_{12}, 
$$
where $g_0$ is the interaction strength. For certain values of $g_0$ and
$\omega$, combined with particular symmetry, one can find 
simple analytical eigenstates. For $g_0=1$ and $\omega=1/4$,
 the lowest triplet of
$P$ symmetry $^3P(sp)$ is given exactly as [19]
$$
\Psi_{exact} = \Psi_0 (1+r_{12}/4),
$$
where the noninteracting solution $\Psi_0$ (\ie $g_0=0$) is as 
usual
$$
\Psi_0=e^{-(r_1^2+r_2^2)\omega/2}(z_1-z_2).
$$
The noninteracting energy for this particular state ($n_1=n_2=l_1=0, l_2=1$) can
 be expressed as
$$E_0=(2n_1+2n_2+l_1+l_2+3)\omega=4\omega=1,
$$ 
while the interacting exact eigenvalue
is 
$$E_{exact}=E_0+1/4.$$
These analytical solutions are sufficiently simple so that we can evaluate 
the nda components for various combinations of Hamiltonians and wave  functions.

a) {\em Noninteracting Hamiltonian and noninteracting 
wave function.} It is straightforward to find out that
for $g_0=0$,  we get 
$$
E_{pot}^{nda} =\frac{7 \omega}{2}=\frac{7}{8},
$$  
and correspondingly
$$
E_{kin}^{nda} =\frac{ \omega}{2}.
$$

b) {\em Interacting Hamiltonian with $g_0=1$ and the exact eigenstate.}
After making transformation to center of mass and relative coordinates,
one can find 
$$
E_{pot}^{nda} =\frac{7 \omega}{2} +\frac{3}{8}\frac{\sqrt{\pi}}{4+3\sqrt{\pi}}
+ \frac{1+\sqrt{\pi}/2}{4+3\sqrt{\pi}} 
$$
and using the exact result above, 
we find
$$
E_{kin}^{nda}= E-E_{pot}^{nda}.
$$

c) {\em Interacting Hamiltonian and noninteracting wave function with the correct node.}
It is interesting to find out the estimation energy considering an approximate
wave function which has the exact node. Let us first consider the noninteracting 
wave function. This will give quite a poor estimate since the potential and kinetic 
energy will be ``unbalanced,'' but it will still be instructive. Taking $\Psi_0$ above, we get
$$
E_{pot}^{nda}=\frac{7 \omega}{2}+\frac{\sqrt{\pi\omega}}{4}
$$
and
$$
E_{kin}^{nda}=\frac{ \omega}{2}.
$$
This provides a clear demonstration that the energy obtained as nda sum is not
necessarily an upper bound since 
$$
E_{kin}^{nda}+E_{pot}^{nda} \cong \frac{9.77 ... \omega}{2} < E_{exact}= \frac{10\omega}{2},
$$
which gives 1.226 ... vs. the exact value 5/4. 
Actually, the error is not very large considering how crude the trial state is.
The dominant error is in the potential part, which comes out lower. 
This is caused by two effects: 
the noninteracting value of the  exponent in the gaussian is not optimal 
and a secondary impact comes also from the absence of the correlation. The kinetic 
energy component is the same as in the noninteracting Hamiltonian, \ie slightly larger
than the exact. This results from the missing exchange hole which affects 
the gradient of the wave function on the nodal surface.
%
Obviously, these ideas should be explored further and such investigations are currently 
in progress. 





{\em Conclusions.} We have introduced the nodal hypersurface and domain averages, dubbed
``nda'',
as a tool for characterization of the nodes of trial wave functions. We have demonstrated 
their properties on a number of few-particle cases and analyzed implications of these results.
For example, we were able to distinguish the nodal differences between degenerate 
states of the same and different symmetries. 
These characteristics enabled us to identify the equivalence of nodes 
in unexpected situations such as between
 noninteracting singlets and triplets. Clearly, the results show interesting potential 
 and deserve further investigation.  The theory can be further
explored with much more powerful developments which will be presented elsewhere.


\begin{acknowledgments}
This work is supported by ARO, DOE and by the NSF grants
DMR-0804549 and OCI-0904794. Discussions with Wei Ku on topics related
to this paper are gratefully acknowledged.
\end{acknowledgments}

\bigskip
\bigskip
{\bf References.} 

[1] D.M. Ceperley and M.H. Kalos, Quantum Many-Body Problems,
in {\sl Monte Carlo Methods
in Statistical Physics}, edited by K.
Binder, pp.145-194, Springer, Berlin, 1979;
K.E. Schmidt and D. M. Ceperley, Monte Carlo techniques for quantum fluids,
solids
and droplets, in {\sl Monte Carlo Methods in Statistical
Physics II}, pp.279-355,
edited by K. Binder, Springer, Berlin, 1984.

[2]  B.L. Hammond, W.A. Lester, Jr., and P.J. Reynolds,
{\sl Monte Carlo Methods in ab initio quantum chemistry}, World Scientific,
Singapore, 1994

[3] J.B. Anderson, Exact quantum chemistry by
Monte Carlo methods, in {\sl
Understanding Chemical Reactivity}, pp. 1-45,
ed. S.R. Langhoff, Kluwer, Dordrecht,
1995.

[4] D. M. Ceperley and L. Mitas, Quantum Monte Carlo methods in
 chemistry, Adv. Chem. Phys. Vol. {\bf XCIII},
pp. 1-38,
Ed. by I. Prigogine and S. A. Rice,
Wiley, New York, 1996.

[5] W.M.C. Foulkes, L. Mitas, R.J. Needs and G. Rajagopal,
  Quantum Monte Carlo for Solids, Rev. Mod. Phys.
{\bf 73}, pp. 33-83 (2001)

[6] K. E. Schmidt and J.W. Moskowitz, Correlated Monte Carlo wave functions for the atoms He through Ne, J. Chem. Phys. {\bf 93},
 4172 (1990)
 
[7] J. Kolorenc, L. Mitas,  Quantum Monte Carlo calculations of
structural properties of FeO solid under pressure, Phys. Rev. Lett.
{\bf 101}, 185502 (2008)

[8] M. Bajdich, L.K. Wagner, G. Drobny, L. Mitas, K. E. Schmidt,
Pfaffian wave functions for electronic structure quantum Monte
Carlo, Phys. Rev. Lett. {\bf 96}, 130201 (2006); see also
Phys. Rev. B {\bf 77}, 115112 (2008)

[9] L. Mitas, Structure of  fermion nodes and nodal cells,
Phys. Rev. Lett. {\bf 96}, 240402 (2006); 
M. Bajdich, L. Mitas, G. Drobny, and L. K. Wagner,  Approximate 
and exact nodes of fermionic wavefunctions: coordinate transformations 
and topologies, Phys. Rev. B {\bf 72}, 075131 (2005)

[10] J. Kolorenc and L. Mitas, Applications of quantum Monte Carlo in condensed systems, 
Rep. Prog. Phys.  {\bf 74}, 026502 (2011).

[11] J.C. Grossman, M. Rohlfing, L. Mitas, S.G. Louie, and M.L. Cohen,
High accuracy many-body calculational approaches for excitations
in molecules, Phys. Rev. Lett. {\bf 86}, 472 (2001)

[12] L.K. Wagner, M. Bajdich, L. Mitas, QWalk: quantum Monte Carlo
code for electronic structure, J. Comput. Phys., {\bf 228}, 3390 (2009).

[13] M. Bajdich, L. Mitas,  Electronic structure quantum Monte Carlo, Acta Physica Slovaca, {\bf 59}, 
81-168 (2009);
 L. Mitas, Electronic structure by Quantum Monte Carlo: atoms,
molecules and solids, Comp. Phys. Commun., {\bf 97}, 107 (1996)

[14] J.B. Anderson,
Quantum chemistry by random walk, J. Chem. Phys. {\bf 65}, 4121 (1976)

[15]  D. M. Ceperley and B.J. Alder, Quantum Monte Carlo,
Science,
{\bf 231}, 555 (1986); 

[16] F. Kr\"uger and J. Zaanen, Fermionic quantum criticality and 
the fractal nodal surface,
Phys. Rev. B {\bf 78}, 035104 (2008)

[17] C. J. Umrigar, K. G. Wilson, and J. W. Wilkins, Optimized trial 
wave functions for quantum Monte Carlo calculations, Phys. Rev. Lett. {\bf 60}, 1719 (1988).

[18] [2] D. Bressanini, D. M. Ceperley, P. J. Reynolds, in: W. A. Lester (Ed.), Recent Advances in Quantum Monte Carlo Methods, World Scientific, Singapore, 2002, Pt. II, p. 1.

[19] M. Taut, Two electrons in an external oscillator potential: Particular analytic solutions of a Coulomb	correlation problem, Phys. Rev. A {\bf 48},  3561 (1993)


\end{document}